\documentclass[conference]{IEEEtran}

\usepackage{cite}
\usepackage{amsmath,amssymb,amsfonts}
\usepackage{algorithm}
\usepackage{algorithmic}
\usepackage{graphicx}
\usepackage{textcomp}
\usepackage{xcolor}
\usepackage{colortbl}
\usepackage{booktabs}
\usepackage{multirow}
\usepackage{url}
\usepackage{array}

\def\BibTeX{{\rm B\kern-.05em{\sc i\kern-.025em b}\kern-.08em
    T\kern-.1667em\lower.7ex\hbox{E}\kern-.125emX}}

\begin{document}

% Table I and Figure 1 are the only two-column-spanning floats in this
% paper; capping dbltopnum at 1 stops them ever stacking at the top of
% the same page, without forcing a hard page break or leaving either one
% stranded alone on an otherwise-empty page. No effect on any other
% float, since none of the rest are table*/figure*.
\setcounter{dbltopnumber}{1}

% -----------------------------------------------------------------------
% Double-blind: no author names, affiliations, emails, funding
% acknowledgment, or identifying links appear below. Do not fill this in
% until camera-ready.
% -----------------------------------------------------------------------
\title{Kafila: Serving Large Language Models on a Trusted Set of Heterogeneous Commodity Machines}

\author{
\IEEEauthorblockN{Murtaza Rangwala, Richard O. Sinnott and Rajkumar Buyya}
\IEEEauthorblockA{
\textit{Quantum Cloud Computing and Distributed Systems (qCLOUDS) Lab}\\
\textit{School of Computing and Information Systems} \\
\textit{The University of Melbourne, Australia} \\
Email: mrangwala@student.unimelb.edu.au, \{rsinnott, rbuyya\}@unimelb.edu.au}
}
\maketitle

\begin{abstract}
Between them, the members of a research group or a circle of friends own
several consumer computers, none large enough to run a capable
large language model. Existing systems pool such capacity across open swarms anyone
may join, which a group admitting only trusted machines cannot use. Bounding membership removes what they depend on: a swarm holds each part
of the model on several peers and routes around a slow one. A
bounded session must use every device it admits. Its
pipeline advances at the pace of whichever device received a share it
cannot serve quickly, so the division has to be right before serving
begins. We propose Kafila, whose protocol assembles a ring from behind NATs,
preferring direct paths and relaying where traversal fails, while its planner measures each device's memory bandwidth,
capacity and reachability, divides the model exactly for a fixed ring
order, and places the head, which holds the embedding and output
projection, together with that division rather than beforehand. On machines with different capabilities across three fleets, from a shared LAN to
five devices spanning two continents, Kafila shortens the slowest
pipeline stage by up to $5.2\times$ against the even split of pipeline
parallelism, as in GPipe, and up to $3.0\times$ against the
memory-proportional split of personal-device inference, as in exo, keeps
75 to 87 per cent of
the committed hardware doing work where those divisions fall below half,
and serves a model no uniform split can place on the fleet at all. What
that is worth to a user depends on how much of a token is computation
rather than network. Where the members share a network the same division
returns $1.56\times$ the throughput of a uniform split and $1.25\times$
of a memory-proportional one, and under four concurrent users that lead
compounds to $3.2\times$ rather than fading, each user served at almost the
rate of one.
\end{abstract}


% =========================================================================
\section{Introduction}
\label{sec:intro}

Between them, the members of a research group or a circle of friends own
several computers capable of running neural networks: laptops, desktops
with discrete graphics cards, the odd workstation.
None is large enough on its own to run a capable large language model (LLM). This is a familiar class of
problem in pervasive computing: the useful resource is the set of
devices a group already has, not any one of them, and the task is to
coordinate them rather than to acquire more.

Our motivating application is a model that answers questions about a
person's own documents, messages, and health
records~\cite{dehghani2025legal,he2025healthcare}, the content people
are least willing to send elsewhere. The most capable models are reached
only through APIs~\cite{menlo2025llmmarket}, so using them means sending
that content to infrastructure the user cannot audit. Open-weight models
close the capability gap~\cite{qwen2025technical} but not the hardware
gap: a competitive model does not fit in the memory of any single device
the group owns~\cite{sheng2023flexgen}, though it may fit in all of them
together. Turning that aggregate into a usable model server is a
device-to-device coordination problem.

One line of work closes the hardware gap by pooling capacity across the
Internet, splitting a model's layers across an open swarm of
volunteers~\cite{borzunov2023petals,ryabinin2023swarm,tong2025parallax}.
Such a pool assembles many small machines into an effectively larger
one, and can grow without bound because anyone may join. The property that makes such a pool large, however, also makes it
unsuitable here. A user who wants to run a model only on machines they
trust cannot use an open swarm. Restricting membership is not a
configuration choice in these systems; it removes the assumption they
are built on. Two consequences follow, the second of which is the
subject of this paper.

The first concerns who holds a user's activations. In an open swarm they
travel to whichever volunteer holds the corresponding block, and a
server holding them can reconstruct the input with high fidelity, even
against common perturbation
defenses~\cite{fan2026actinv,erdogan2022unsplit}. Confining the group to
invited participants keeps every block with an invited peer, so
activations reach no unknown volunteer, and a relayed edge that must cross
the rendezvous crosses it end-to-end encrypted, as ciphertext the rendezvous
cannot read; the protocol prefers direct paths all the same
(Section~\ref{sec:protocol-relay}), and Section~\ref{sec:eval-protocol}
measures how often it relays per fleet. We treat the
block-level guarantee as a premise of the design rather than a
contribution.

The second consequence is that a bounded group is harder to run a model
on quickly. An open swarm tolerates a slow participant because it has
redundancy: many peers hold each block, so a client can be routed around
a poor one, which is what Parallax's scheduler
does~\cite{tong2025parallax}. A bounded group of five invited devices has no
such freedom. Every admitted device must be used however slow, and a
pipeline advances at the pace of its slowest stage. Existing systems do
not plan the forward pass against measured per-device capability. That
slowest stage is therefore whichever device received an equal share of
the model rather than the share it could sustain. Throughput suffers
even where the aggregate hardware would allow better. For instance, Petals reports
roughly one decoding step per second for a 176-billion-parameter
model~\cite{borzunov2023petals}. A bounded group therefore does not need
a better route through a large pool. It needs a division of the model
that suits the devices it actually has.

We call such a bounded group a \emph{session}: it is assembled on
demand to serve one model and dissolves when its last member leaves.
Membership is defined by relationship rather than physical proximity:
at one extreme, a session is a single household's devices, necessarily
co-located; at the other, it is a research group's workstations, or a
group of friends' machines, spread across cities or continents.
Scoping by relationship rather than by proximity, however, is what
makes a session hard to realize, and two difficulties follow, both of
which this paper addresses. First, the devices sit behind independent
home or institutional Network Address Translators (NATs), and cannot
reach one another until they are told how. A pipeline needs every
adjacent pair in the ring to connect. One unreachable pair therefore
does not merely slow a session down; it stops the session from forming
at all. Second, the devices differ by an order of magnitude in memory and
in memory bandwidth, and no redundancy exists to absorb that
difference. The division of the model therefore decides both how fast a
session runs and whether it runs at all. We present Kafila, a session
protocol and planner for exactly this setting. It makes the following
contributions:\smallskip

\begin{itemize}
\item \textbf{A session protocol} that forms a pipeline entirely from
  behind NAT, with no router configuration, using an always-on
  rendezvous service for signaling only. It prefers hole-punched direct
  paths and relays only where traversal fails.
\item \textbf{A planner} that decides block assignment and head
  placement from each device's measured memory bandwidth and capacity,
  over a ring ordered from measured reachability. Placement is solved exactly for a
  fixed ring order by chain-on-chain partition, head placement jointly
  with that division rather than beforehand, and a session for which no
  assignment fits is declined.
\item \textbf{An evaluation} across three fleets, from a shared LAN to
  five devices on two continents, on models up to 32 billion parameters.
  Dividing against measured capability, the planner shortens the slowest
  pipeline stage by up to $5.2\times$ where capability-blind divisions leave
  much of the hardware idle, and serves a model no uniform split can place
  at all.\smallskip
\end{itemize}

The rest of the paper is organized as follows.
Section~\ref{sec:related} situates Kafila relative to the open-swarm and
pipeline-parallelism literature. Section~\ref{sec:model} defines the
session model. Section~\ref{sec:protocol} describes the protocol that
forms a session from behind NAT. Section~\ref{sec:planner} presents the
planner. Section~\ref{sec:impl} describes the implementation, the
testbed, and the NAT emulation the protocol results rest on.
Section~\ref{sec:eval} reports the evaluation.
Section~\ref{sec:conclusion} concludes and states what remains open.

% =========================================================================
\section{Related Work}
\label{sec:related}

\begin{table*}[t]
\caption{Comparison with related systems}
\centering
\footnotesize
\setlength{\arrayrulewidth}{0.4pt}
\setlength{\tabcolsep}{4pt}
\begin{tabular}{
  >{\raggedright\arraybackslash}p{2.55cm}
  >{\raggedright\arraybackslash}p{2.10cm}
  >{\raggedright\arraybackslash}p{2.75cm}
  >{\raggedright\arraybackslash}p{2.30cm}
  >{\raggedright\arraybackslash}p{2.95cm}
  >{\centering\arraybackslash}p{1.45cm}
  >{\centering\arraybackslash}p{1.35cm}
}
\toprule
System & Workload & A slow or absent participant & Connectivity graph & Per-stage cost obtained from & Ordering by reachability & Head placed with the division \\
\midrule
\rowcolor{black!6}
Petals~\cite{borzunov2023petals} & LLM inference & Routed around at request time & Complete & Throughput measured at runtime & $\times$ & $\times$ \\
SWARM~\cite{ryabinin2023swarm} / Parallax~\cite{tong2025parallax} & LLM inference & Rescheduled at runtime & Complete, unequal bandwidth & Throughput measured at runtime & $\times$ & $\times$ \\
\rowcolor{black!6}
GPipe~\cite{huang2019gpipe} & Training & Does not arise; deployment is provisioned & Complete, uniform speed & Layer count & $\times$ & $\times$ \\
exo~\cite{exo} & LLM inference & Slows the ring; not addressed & Complete, on one network & Device memory & $\times$ & $\times$ \\
\rowcolor{black!6}
PipeEdge~\cite{hu2022pipeedge} / PipePar~\cite{zhang2023pipepar} & Encoder inference; training & Does not arise; deployment is provisioned & Complete, unequal speed & A profile taken with the weights resident & $\times$ & $\times$ \\
\textbf{Kafila (this work)} & LLM inference & Cannot be avoided, so it is planned for & Incomplete and unequal & A bandwidth probe, before any weight is fetched & $\checkmark$ & $\checkmark$ \\
\bottomrule
\end{tabular}
\label{tab:comparison}
\end{table*}

\subsection{Open-swarm collaborative inference}

Petals~\cite{borzunov2023petals} is closest in spirit. It partitions a
model across volunteer machines and pipelines requests through them,
the same workload structure Kafila targets, but differs in population
and hence in guarantees. Petals' swarm is open and unbounded,
tolerating churn through redundancy. A departing server is routed
around because another elsewhere holds the same block. A session, on
the other hand, has no redundant capacity to route around with. This is not a design decision that could be
reversed. A user cannot both confine a model to machines they trust and
retain a pool large enough to hold each block several times over, since
the number of machines a person has a relationship with is small. The
problem is therefore to make a small, fixed membership functional, not
to make an unbounded population resilient. SWARM
parallelism~\cite{ryabinin2023swarm} and Parallax~\cite{tong2025parallax}
extend the open-swarm model to training and to explicitly
heterogeneous, multi-datacenter GPU pools with dynamic scheduling, but
retain the same assumption of an unbounded, changing peer set with no
guarantee over ring adjacency. Both mitigate the resulting performance
variance by routing and scheduling at runtime rather than by planning
in advance. This is the reasonable choice for an open population, which
changes too fast for a plan to remain valid. It is also what a bounded
group cannot borrow. Routing around a poor peer requires a peer to route
to, and a session has none, so the quality of the initial division is
the only lever available.

\subsection{Pipeline parallelism over heterogeneous, imperfectly connected machines}

Pipeline parallelism was developed for the datacenter, and its
partitioning rules are the ones practitioners still inherit.
GPipe~\cite{huang2019gpipe} divides a network into stages of equal layer
count, which is sound when the devices are identical. exo~\cite{exo},
which assembles a ring across phones, laptops and desktops, instead
gives each device a share proportional to its memory, which is sound
when memory and reading speed rank alike. A session guarantees neither.
Its devices differ in speed, and the one with the most memory is often
not the fastest, so both rules hand blocks to a device that cannot serve
them at the rate the division assumes.

PipeDream~\cite{narayanan2019pipedream} instead minimizes the slowest
stage by dynamic programming, which is Kafila's objective, and
PipeEdge~\cite{hu2022pipeedge} and PipePar~\cite{zhang2023pipepar}
extend that to unequal devices. All three assume a complete graph and a
deployment administered as one system. A session has neither: its pairs
can communicate only sometimes, and which pairs is unknown until
measured. Two departures follow. Ring order is chosen from measured
reachability rather than given. And stages are not interchangeable,
since the head holds the embedding and the output projection, so it
costs more per token and holds fewer blocks at equal memory, an
asymmetry Kafila carries into both the per-device cost function and the
per-device block limit so that head placement is solved together with
the division.

The decomposition rests on Pinar et al.~\cite{pinar2008oned}, who show
that \emph{chain-on-chain} partitioning, a contiguous chain divided
across heterogeneous processors whose order is fixed, is exactly
solvable in polynomial time, while choosing that order as part of the
same problem is NP-complete. Kafila fixes the ring order first, from
measured reachability, then solves block placement exactly.

\subsection{NAT traversal}

Hole punching establishes a path through a NAT by having both peers
send outbound packets, which the middlebox then learns to forward
responses through. Ford et al.~\cite{ford2005p2p} documented and
analyzed it for UDP and TCP. Its building blocks are standardized in
STUN~\cite{rfc5389} and ICE~\cite{rfc8445}, and the NAT behavior
taxonomy Kafila reasons about follows RFC~4787~\cite{rfc4787}.
Trautwein et al.~\cite{trautwein2026nat} measured hole punching across
more than 85{,}000 IPFS networks. Roughly 70\% of attempts succeed once
relay reservation and address discovery complete. That is an
appropriate anchor for an open population of arbitrary peers. For a
session's smaller and more idiosyncratic membership it is an assumption
to verify rather than to inherit.

\subsection{Summary}

Table~\ref{tab:comparison} sets these systems against the single
question that separates them: what happens when a participant is slow
or absent. A swarm reschedules around it; a provisioned deployment never
meets it. A session has neither a spare peer nor an administrator, so
the division of the model is decided once, in advance, from what the
devices happen to be and where they happen to sit. That is why Kafila
must get the division right before serving begins, with no runtime
recovery to fall back on, and why it must order members by measured
reachability, a step a cluster never needs.

% =========================================================================
\section{The Kafila Session Model}
\label{sec:model}

A \emph{session} is a bounded set of devices, assembled on demand, that
together serve one model for as long as its members remain. Three roles
participate. The \emph{rendezvous} is one always-on component,
reachable outbound by every device, that carries session registration
and the signaling required for NAT traversal. It holds no model weights
and performs no inference, so it needs neither an accelerator nor
substantial compute, and a small cloud instance is sufficient. It
carries activations only on an explicit, last-resort fallback, and then only
as ciphertext (more on this in Section~\ref{sec:protocol-relay}); even then
its cost is bandwidth rather than computation. The \emph{host} creates the
session and selects the model. Every other participant is a
\emph{member}, joining with a session code and reporting its measured
capacity.

\begin{figure*}[t]
\centering
\includegraphics[width=0.78\textwidth]{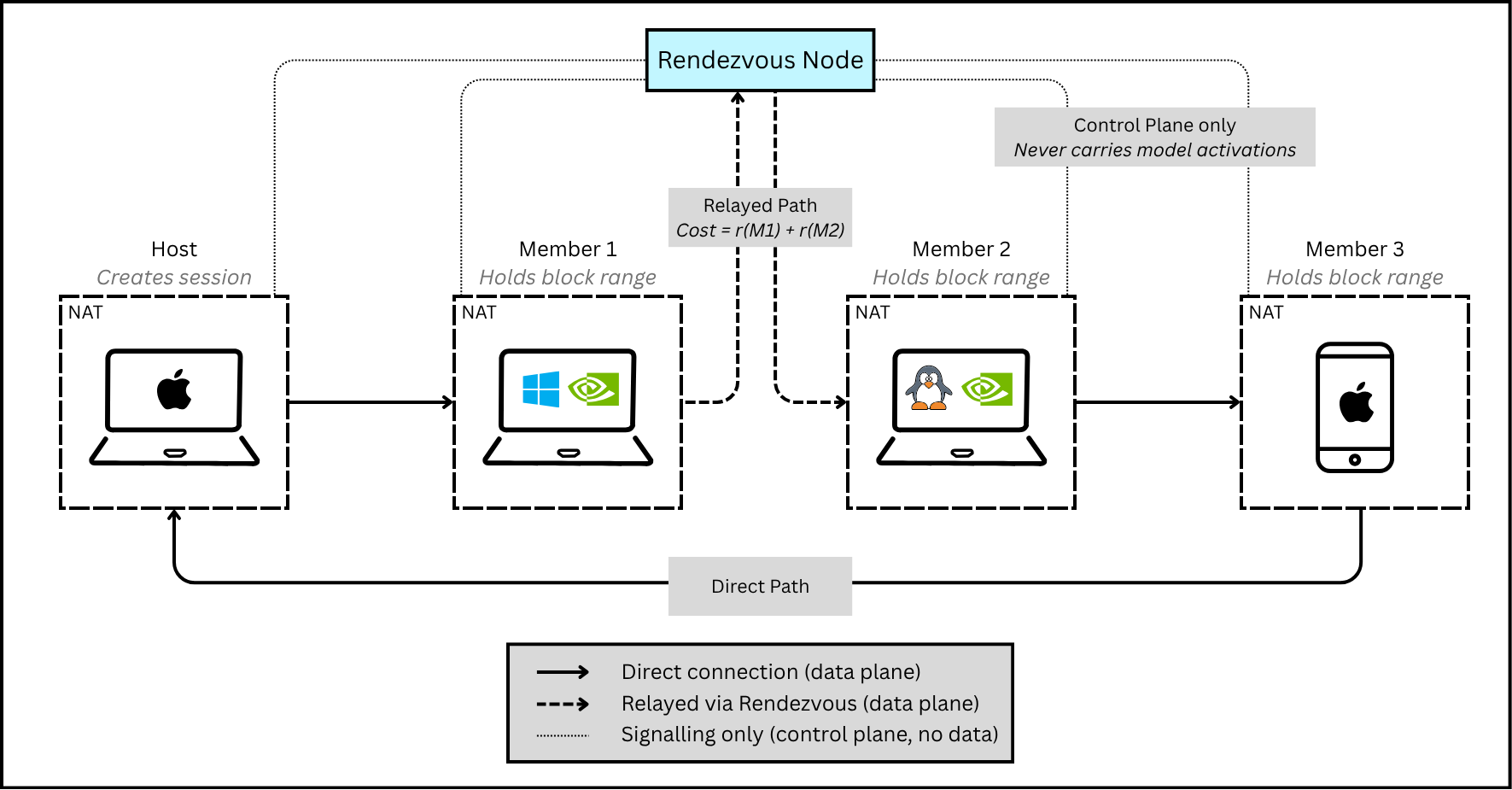}
\caption{A four-member session model, arranged as a ring from the host
through three members and back.}
\label{fig:architecture}
\end{figure*}

One member holds the \emph{head}: the embedding, the unembedding, and
the sampler, which are not divisible across shards. Which member holds it
is decided by the planner rather than by who opened the session. The
head's components are resident in addition to its blocks, so the choice
of head and the division of blocks constrain each other, and
Section~\ref{sec:planner-blocks} solves them together.

A session accepts prompts at every member, not only at the head,
though only the head can actually answer one. Each member serves the
same interface locally and forwards it, so a person uses the session
from the device in front of them without knowing which machine is the
head. This is the reciprocity the setting calls for. The laptop that
contributes a shard is also where its owner types. It also means a
session may be asked for several answers at once, by different people,
which the planner's cache reservation must account for.

Membership defines the trust boundary. The session code is a bearer
token, so any holder can join and will observe the activations that
cross the shard it is assigned. This is the whole of the access-control model.
A session is scoped to whoever the host shares the code with, which is
the boundary a household, a laboratory, or a group of collaborators
already maintains.

The members are arranged in a \emph{ring}. A token passes from the
head through each member's block range in turn and returns to the head.
The return leg is what makes it a ring rather than a line: the head owns
the unembedding, so the last member's output has to come back before a
token can be produced. Within the ring, each member's compute overlaps
the next member's transfer, as in prior pipeline-parallel systems. What
a session adds is that the cost of an edge depends on whether its two
members can reach each other directly, which is not known before the
session is formed. Choosing the ring order, and the block assignment
within it, is the planner's responsibility.

A \textit{session} model with four members is illustrated in Figure~\ref{fig:architecture}. Every
member holds a long-lived signaling connection to the rendezvous, used
only for registration and NAT-traversal coordination. The ring carrying
the data plane is separate, and its edges are direct wherever traversal
succeeds and relayed through the rendezvous only where it does not.

% =========================================================================
\section{Protocol: Forming a Session From Behind NAT}
\label{sec:protocol}

A naive approach has the host listen on a port and publish an address
for members to connect to, and has every ring member listen for its
predecessor. Both directions require accepting an unsolicited inbound
connection, which a device behind NAT cannot do without manual router
configuration. That assumption does not hold for most users. The
protocol's governing constraint is therefore that no participant is ever
required to accept an unsolicited inbound connection, or to configure a
router. The rest of this section follows from that constraint.

\subsection{Bootstrap and reachability discovery}

The host opens a long-lived \emph{outbound} connection to the rendezvous
and registers a model identifier, an expected member count, and its own
capability, receiving a short session code in return. A member connects
outbound and presents the code. The rendezvous introduces the two,
exchanging each side's candidate addresses.

Every member's reachability is classified before the ring is planned.
Writing $m$ for the number of members, probing every pair directly is an
$O(m^2)$ network operation. It is also unnecessary. Whether a given pair
can connect is largely a function of each node's own NAT
behavior~\cite{rfc4787}: how it maps an internal port to an external one
across different destinations, and whether it accepts a packet from a
host it has not itself sent to. Both are measurable without involving
the other party. Kafila therefore classifies each node once, against the
rendezvous and STUN servers, at $O(m)$ probes; predicts every pair's
feasibility from those measurements, which is $O(m^2)$ arithmetic and no
packets; and then verifies only the edges the chosen
plan uses, at $O(m)$ probes. The quadratic step is a table lookup, so
network cost is linear in membership size at both ends.

Prediction is a prior, not a guarantee, and both directions of error are
tolerated rather than corrected. An edge predicted direct that fails
verification falls back to relaying, and the ring order computed from
the prediction is kept. An edge predicted relayed that connects directly is
cheaper than the plan assumed, and the session runs faster on it than
predicted. Section~\ref{sec:eval-protocol} reports how often each
occurred against real equipment.

\subsection{Traversal and relay fallback}
\label{sec:protocol-relay}

The data plane runs over QUIC~\cite{rfc9000}. Using UDP avoids simultaneous-open TCP
traversal, which succeeds less often than its UDP
equivalent~\cite{ford2005p2p}. Kafila attempts a hole-punched direct
path for every edge first, and relays through the rendezvous only where
that fails.

The preference order is first a latency decision. A direct edge costs
the round trip between its two members. A relayed edge costs the sum of
both members' round trips to the rendezvous, because every frame travels
to it and back out again. A ring pays that difference on every edge it
relays, for every token of every request, so it compounds rather than
being amortized. The planner prices the two cases in the same unit and orders the ring on
the result.

Confidentiality does not turn on the path. The rendezvous mediates
signaling, so it knows the session's membership and the members' addresses,
but each edge's QUIC session is encrypted end-to-end by
TLS~1.3~\cite{rfc9001, rfc8446}, whose keys the rendezvous never holds. Where an edge relays, the rendezvous forwards the
members' encrypted datagrams rather than terminating their session, and so
sees only ciphertext: it learns how much two members exchange, never the
tensors they carry, and a direct edge keeps even that off it. Preferring
direct paths narrows what the rendezvous observes to the signaling it already
has, and removes the dependency on it, rather than protecting content the
encryption protects on every path.
Section~\ref{sec:eval-protocol} reports the fraction of edges that stayed
direct on real consumer connections.

\section{Planner: Cost, Placement, Ordering, and Admission}
\label{sec:planner}

Decoding at batch size one, producing one token at a time for a
single in-flight request, is memory-bandwidth-bound~\cite{pope2023efficiently},
and a ring incurs
its network cost on every token: a request's per-token circuit time is
\begin{equation}
T_{\mathrm{circuit}} = \sum_i c_i + \sum_{(i,j)\in\mathrm{ring}} \ell_{ij},
\label{eq:circuit}
\end{equation}
the sum of each member's compute stage $c_i$ and every ring edge's
latency $\ell_{ij}$. Two properties of \eqref{eq:circuit} pull in
opposite directions. The compute term is invariant to ring size:
splitting a fixed forward pass across more members subdivides the same
total work rather than adding to it. The latency term grows linearly
with ring size, because each additional member introduces another hop
paid on every token. The planner's task is to choose the block
assignment and ring order that minimize \eqref{eq:circuit} for a given
membership, and to determine when no choice is feasible.

\subsection{Cost model}
\label{sec:planner-cost}

Each stage's compute cost $c_i$ is derived rather than fitted, so that
it can be checked against measurement rather than tuned to it. At batch
size one a stage reads every weight it holds exactly once, and performs
two floating-point operations per weight, so memory bandwidth rather
than arithmetic throughput dominates. This gives
\begin{equation}
c_i = o_i + n_i \cdot \frac{\beta_{\mathrm{block}}}{\bar{b}_i} +
      \mathbb{1}[i = \mathrm{head}] \cdot \frac{\beta_{\mathrm{head}}}{\bar{b}_i},
\label{eq:stage-cost}
\end{equation}
where $\bar{b}_i$ is device $i$'s measured memory bandwidth and $n_i$
the number of blocks it is assigned. The byte sizes
$\beta_{\mathrm{block}}$ and $\beta_{\mathrm{head}}$ are properties of
the model, identical for every device: one transformer block, and the
head's embedding and unembedding matrices. The term $o_i$ is a fixed
per-token dispatch overhead, measured independently of role, and
$\mathbb{1}[i=\mathrm{head}]$ is $1$ for the head and $0$ otherwise.
Every term is measured before a weight is loaded. A short device-side
probe supplies $\bar{b}_i$ and $o_i$; the checkpoint's manifest
supplies the two byte sizes. No term is fitted to an observed stage time, so the model can be tested
against stage times it never saw.

Two things follow for head placement, and they are different. The
first is cost, since the head pays $\beta_{\mathrm{head}}/\bar{b}_i$ on
every token, so where that lands matters, though only to the extent
that devices differ in $o_i$ as well, since every other term scales
with bandwidth alike. The second is capacity, since the head's
embedding and unembedding are resident alongside its blocks, so a
device holding the head can hold fewer blocks than the same device
could otherwise. The
first makes head placement worth optimizing. The second makes it a
constraint, and Section~\ref{sec:planner-blocks} solves the two
together.

\subsection{Ring ordering}
\label{sec:planner-order}

Ordering is chosen over a connectivity graph that is \emph{incomplete in
addition to being unequal}: an edge may be direct, relayed at several
times the cost, or, before verification, merely predicted either way.
The planner's first pass is a closed-form heuristic over each member's
NAT behavior class. Following RFC~4787~\cite{rfc4787}, members are
permissive, meaning endpoint-independent in both mapping and filtering;
restrictive, meaning endpoint-dependent mapping; or ordinary in between.
A ring edge can be direct only if at least one endpoint is permissive,
so an all-direct ring exists if and only if the permissive members are
at least as many as the restrictive ones, and alternating the two around
the anchor achieves one whenever it is achievable. This is a counting
argument rather than a search, and stays linear in membership size.

Counting relayed edges is the right objective only where direct edges
are uniformly cheap and relayed ones uniformly expensive, and neither
holds once a session spans more than one site. On our testbed members
measured median round trips of 79 to 115\,ms to the rendezvous, so a
relayed edge cost between 158 and 230\,ms depending on which two members
it joined. A rule that counts treats all of them as one quantity.

Where every member has measured its own round-trip time to the
rendezvous, the planner replaces counting with a model that prices both
kinds of edge in the same unit. Writing $r_X$ for member $X$'s measured
round trip, a relayed edge $A \leftrightarrow B$ costs $r_A + r_B$,
which is not an estimate but the literal path through the rendezvous. A
direct edge is bounded below by the triangle inequality at
$|r_A - r_B|$. Ring cost is the sum of both kinds around the cycle. The
planner searches orderings exhaustively for the cheapest, holding one
member's position fixed to remove the ring's rotational symmetry.

One case is degenerate. When every edge in a ring must relay,
$\sum_{\mathrm{edges}} (r_A + r_B) = 2\sum_{\mathrm{members}} r_i$ for
\emph{any} arrangement of the same members, so every ordering scores
identically and the model is silent exactly where it is needed most: it
can place two members who share a network at opposite ends of the ring.
Kafila breaks such ties on ring \emph{separation}, the same
$|r_A - r_B|$ term summed around the cycle, which prefers orderings that
place members of similar rendezvous distance next to each other. A
predicted-relayed edge sometimes connects anyway, which
Section~\ref{sec:eval-protocol} observes in two of nine class pairs and on
the intercontinental fleet, and the arrangement that gains when it does is
the one whose near members are already neighbors; where the prediction
holds instead, adjacency costs nothing. This does not resolve the model's
blindness between two equally relay-bound arrangements, but it stops the
model choosing the worse of them.

\subsection{Block placement}
\label{sec:planner-blocks}

Fixing the ring order first turns block placement into a
\emph{chain-on-chain} partition: a contiguous chain of transformer
blocks divided across processors of known, unequal speed, in a fixed
processor order. Pinar et al.~\cite{pinar2008oned} show this is solvable
exactly in polynomial time, in contrast to the NP-complete problem of
choosing order and placement jointly.

The planner solves it with a parametric search, given as
Algorithm~\ref{alg:partition} (The \textsc{Partition} algorithm). The search is stated over two per-device
functions derived from \eqref{eq:stage-cost}: $\mathrm{fixed}(i) = o_i$,
plus $\beta_{\mathrm{head}}/\bar b_i$ if $i$ is the head, and
$\mathrm{per}(i) = \beta_{\mathrm{block}}/\bar b_i$. A third
function, $\mathrm{lim}(i)$, is device $i$'s block limit at the
session's declared context and concurrency. It is smaller when $i$ is
the head, because the head's resident state includes the embedding, and
this is where head placement enters as a constraint rather than as a
cost.

A device's stage cost is affine in the number of blocks it holds, so
the bottleneck of any feasible division equals
$\mathrm{fixed}(i) + n\cdot\mathrm{per}(i)$ for some device $i$ and some
integer $n$. The optimum is therefore one of at most $O(k \cdot m)$
candidate values, for $k$ blocks and $m$ devices, rather than an
arbitrary real number (Algorithm~\ref{alg:partition}, line~1). The
search sorts these candidates and binary-searches them for the smallest
that is \emph{feasible} (line~2).

The \textsc{Feasible} subroutine tests a candidate bound in one left-to-right pass,
assigning each device the largest block count it can hold within the
bound, subject to $\mathrm{lim}(i)$ and to reserving one block for every
device still to come. The pass is optimal by an exchange argument: the
chain is contiguous and the order fixed, so any block a device declines
falls to a later one and cannot lower the bottleneck. Feasibility is
monotone in the bound, so binary search over the sorted candidates is
correct. Generating and sorting them dominates, giving $O(km\log km)$
overall. \textsc{Partition} returns $\tau^\star$ alongside the division,
since head placement compares rotations by achieved cost.

\begin{algorithm}[t]
\caption{\textsc{Partition}: exact chain-on-chain block placement for
a fixed device order.}
\label{alg:partition}
\begin{algorithmic}[1]
\REQUIRE $k$ blocks; devices $1..m$ in ring order, device $i$ costing
$\mathrm{fixed}(i) + n\cdot\mathrm{per}(i)$ for $n$ blocks, up to a
limit $\mathrm{lim}(i)$
\ENSURE block counts $n[1..m]$ and their bottleneck $\tau^\star$, or
\textsc{infeasible}
\STATE $C \gets \{\,\mathrm{fixed}(i) + n\cdot\mathrm{per}(i) : 1\le i\le m,\ 1\le n\le \min(k,\mathrm{lim}(i))\,\}$, sorted ascending
\STATE $\tau^\star \gets$ smallest $\tau \in C$ with $\textsc{Feasible}(k, \tau) \neq \textsc{infeasible}$ \hfill\COMMENT{binary search over $C$}
\IF{no such $\tau^\star$}
  \RETURN \textsc{infeasible}
\ENDIF
\RETURN $\textsc{Feasible}(k, \tau^\star)$, $\tau^\star$
\STATE \textbf{function} \textsc{Feasible}($k$, $\tau$):
\STATE\quad $\mathrm{left} \gets k$
\FOR{$i = 1$ \textbf{to} $m$}
  \IF{$\mathrm{fixed}(i) > \tau$}
    \RETURN \textsc{infeasible}
  \ENDIF
  \STATE\quad $n[i] \gets \min\!\big(\lfloor(\tau-\mathrm{fixed}(i))/\mathrm{per}(i)\rfloor,\ \mathrm{lim}(i),\ \mathrm{left}-(m-i)\big)$
  \IF{$n[i] < 1$}
    \RETURN \textsc{infeasible}
  \ENDIF
  \STATE\quad $\mathrm{left} \gets \mathrm{left} - n[i]$
\ENDFOR
\IF{$\mathrm{left} \neq 0$}
  \RETURN \textsc{infeasible}
\ENDIF
\RETURN $n$
\end{algorithmic}
\end{algorithm}

Head placement reuses \textsc{Partition} rather than reopening the
NP-complete joint problem, by the same rotation argument that fixes the
ordering. A ring has no distinguished first device, so rotating the
device sequence leaves every edge where it was, and the ordering already
chosen is unaffected by which device serves as head.
The planner therefore solves the fixed-order partition once per
rotation, $m$ exact solves rather than a search over $m!$ orderings, and
keeps the rotation with the lowest achieved bottleneck, un-rotating its
block counts to the original order.

Two devices of similar bandwidth are close to indifferent as head, since
every non-overhead term in \eqref{eq:stage-cost} scales with bandwidth
alike. The choice then rests on the overhead terms $o_i$, and those are
noisy. The two lowest-overhead devices on our testbed differed in median
$o_i$ by $1.5\,\mu$s, while one of the two varied by $8.4\,\mu$s across
the same sessions. A difference smaller than one device's own spread
would otherwise be enough to move the head.
The planner therefore keeps the anchor as head unless some
rotation improves on it by more than a margin, accepting the alternative
only when $\tau^\star < \tau_{\mathrm{anchor}}(1-\mu)$ with $\mu = 0.05$, chosen against the noise it absorbs rather than against
any particular gain. Overhead accounts for at most 2.2 per cent of a
measured stage across every device and division we ran, and its
run-to-run variation moves a stage by at most 1.1 per cent, so an
apparent five per cent improvement is larger than overhead noise can
produce.

\subsection{Admission}
\label{sec:planner-admission}

Ordering and placement both depend on inputs a device may not have.
A device that failed its bandwidth probe cannot be priced by
\eqref{eq:stage-cost}, and one that failed to time itself against the
rendezvous cannot be ordered by measured cost. Refusing to plan whenever
an input is missing is not an option, since a session is only as well
instrumented as its worst-measured member. The planner instead degrades
through three tiers, each better informed than the one below.

If every member priced itself under the full cost model, the search of
Section~\ref{sec:planner-blocks} runs as described. If some could not,
but every member reported raw arithmetic throughput, the same search
runs with $\mathrm{per}(i)$ set from that ratio and no head term, since
arithmetic throughput does not price the head's extra read; the head
then stays where the session was opened. Only the lowest tier abandons
the search: with nothing measured, blocks are divided in proportion to
reported working set, which is the least informed division and the one
both tiers above improve on.

Admission is checked once, after a division has been produced by
whichever tier applies, against every member's available room at the
session's declared context and concurrency. If the assignment exceeds
what a device can hold, the plan is rejected and the undersized device
is named, rather than reporting only that the session cannot run. When the full cost model has already
tried every division that minimizes the bottleneck and none fits, the
only remaining reason to refuse is memory. The planner refuses, rather
than return a plan that would exhaust a device partway through
serving. This is a feasibility criterion rather than a usability one. It
establishes that a division exists within the members' collective
memory, not that the resulting circuit time is worth serving.

\section{Implementation and Experimental Setup}
\label{sec:impl}

Kafila is implemented as a fork of Ollama~\cite{ollama}, a widely used
local-inference server, retargeted to MLX~\cite{mlx2023}, whose Metal and
CUDA backends let the same session and planner code run unmodified
across Apple silicon and Nvidia hardware. A session's membership mixes
consumer platforms, so no code path is specific to a platform or a
region. Weights are fetched per shard from the model host's safetensors files
by byte range, so a member holding one-tenth of a model downloads one
tenth of it. Shards of the same checkpoint share tensors, so replanning
a division after a partial fetch costs little.

The rendezvous builds without the inference runtime, since it holds no
model and needs no accelerator, and runs as an unprivileged service on a
two-core cloud instance. Every session records predicted and measured
cost, circuit time, and bytes transferred, per stage and per hop, and
each trace separates time-to-first-token into prefill (processing the
prompt) and decode. Each cell is five generations, the first discarded
as cold, and reported data is the median of the rest. All the data in this
paper is captured using that
instrumentation.

\subsection{Testbed and NAT emulation}
\label{sec:impl-testbed}

The three fleets hold the members roughly fixed and vary how far apart
they sit (Table~\ref{tab:testbed}), so that the network overhead, the
fraction of each token spent on the network rather than computing, runs from
12 per cent on the shared LAN to 94 across continents. Every fleet mixes consumer graphics cards, consumer-sized slices of
larger cards, an institutional GPU, and an Apple laptop, and on every one
speed and memory rank differently: the LAN fleet's A40 holds the least memory yet reads faster
than the laptop, and the laptop holds middling memory while reading an
order of magnitude slower than the fastest card. No share assigned by
memory alone matches what a device can serve. Across the fleets bandwidth
spans an $11\times$ range and memory a $4\times$ range. Home uplinks are
not reproduced, which affects fetch time rather than the served rate we
report, and intermittency and churn are outside this paper's scope.

\begin{table}[htbp]
\caption{The three evaluation fleets}
\centering
\footnotesize
\setlength{\tabcolsep}{4pt}
\begin{tabular}{
  >{\raggedright\arraybackslash}p{1.50cm}
  >{\raggedright\arraybackslash}p{1.4cm}
  >{\centering\arraybackslash}p{0.5cm}
  >{\centering\arraybackslash}p{1.35cm}
  >{\raggedright\arraybackslash}p{2cm}
}
\toprule
Fleet & Device & Mem & Bw (GB/s) & Site \\
\midrule
LAN & L40S & 24 & 512--529 & Melbourne, AU \\
(4 dev) & L40 & 24 & 367--453 & Melbourne, AU \\
 & A40 & 12 & 377--439 & Melbourne, AU \\
 & M3 Pro & 18 & 89--124 & Melbourne, AU \\
\midrule
US Central & A100 & 80 & 915--959 & Des Moines, US \\
(5 dev) & RTX 5090 & 32 & 849--1047 & Chicago, US \\
 & RTX 3090 & 24 & 673--714 & Marion, US \\
 & RTX A6000 & 48 & 393--436 & Kansas City, US \\
 & RTX A6000 & 48 & 400--480 & Kansas City, US \\
\midrule
Inter- & RTX 5090 & 32 & 1303--1405 & Arad, RO \\
continental & A100 & 40 & 812--949 & Prague, CZ \\
(5 dev) & RTX 3090 & 24 & 625--704 & Teplice, CZ \\
 & L40S & 24 & 505--526 & Melbourne, AU \\
 & M3 Pro & 18 & 89--127 & Melbourne, AU \\
\bottomrule
\end{tabular}
\label{tab:testbed}
\end{table}

The three models served are Qwen3-8b, Qwen3-14b and
Qwen3-32b~\cite{qwen2025technical}, each generation decoded greedily at a
fixed seed so allocations are compared on identical output. The class
pairs that bear most on the ordering rule are the ones a real network is
least likely to offer, so those results are obtained under Tailscale's
\texttt{natlab}~\cite{natlab}, which models the RFC~4787 matrix directly.
Only the NATs are emulated: the candidate exchange, the hole punch, the
QUIC handshake, the reachability classifier, and the rendezvous carrying
signaling are the deployed implementation, running unmodified over
emulated links.

% =========================================================================
\section{Performance Evaluation}
\label{sec:eval}

The protocol and the planner rest on different instruments, so we
evaluate them apart. The protocol's claim is about NAT behavior, which we
exercise on real consumer connections and, for the classes a real
network seldom offers, under emulation. The planner's claim is about
heterogeneous compute, which needs unequal real machines rather than a
simulated cost model, so it is measured on the three fleets of
Table~\ref{tab:testbed}.

We compare Kafila against two existing algorithms~\cite{huang2019gpipe, exo}. \textsc{Uniform} gives every member the
same number of blocks, which is GPipe's rule~\cite{huang2019gpipe}.
\textsc{Memory-proportional} gives each a share proportional to its
working set, exo's default for rings of personal
devices~\cite{exo}. PipeEdge~\cite{hu2022pipeedge} and
PipePar~\cite{zhang2023pipepar} price a stage from a profile taken with
the weights already resident, which a session cannot obtain before it has
decided who holds what. Each cell is five independent sessions; the
network is drawn afresh in each, so throughput carries its five-session
half-range while the compute terms, taken on each device's own clock, do
not.

\subsection{Forming a session across real NATs}
\label{sec:eval-protocol}

Real networks separate the cases the emulator cannot. On the US Central
fleet the five members sat behind five different public networks, and
traversal punched every ring edge direct over public addresses at a median
of 1.29\,s, so no activation ever left the served ring. The four LAN members instead sat on one private network, and their edges
connected over local addresses at 1.26\,s; the reachability rule read all
four as restrictive and predicted relays, but the verdict cost nothing
because no edge needed the rendezvous to begin with. Across continents the
fallback carries the load: on the intercontinental fleet only the two
co-located members reached each other directly, at 2.24\,s, and the other
four ring edges relayed at a median of 20.8\,s, so roughly four fifths of
each token's activation bytes crossed the rendezvous as ciphertext. A ring planned around
a relay that proves unnecessary costs nothing at run time, whereas the
reverse would strand it on a path that does not exist, so the rule errs
only in the safe direction. In every session the ordering search returned
an arrangement optimal in predicted relayed edges, verified against an
exhaustive check.

\providecommand{\tmatch}[1]{\textcolor{green!60!black}{#1}}
\providecommand{\tmiss}[1]{\textcolor{red!80!black}{#1}}
\begin{table}[t]
\caption{NAT class-pair traversal outcomes}
\label{tab:classpairs}
\centering
\footnotesize
\setlength{\tabcolsep}{5pt}
\begin{tabular}{ll cc}
\toprule
Initiator & Responder & Rule predicts & Traversal achieves \\
\midrule
permissive & permissive & \tmatch{direct} & \tmatch{direct} \\
permissive & ordinary & \tmatch{direct} & \tmatch{direct} \\
permissive & restrictive & \tmatch{direct} & \tmatch{direct} \\
ordinary & permissive & \tmatch{direct} & \tmatch{direct} \\
ordinary & ordinary & \tmatch{direct} & \tmatch{direct} \\
ordinary & restrictive & \tmiss{relay} & \tmiss{direct} \\
restrictive & permissive & \tmatch{direct} & \tmatch{direct} \\
restrictive & ordinary & \tmiss{relay} & \tmiss{direct} \\
restrictive & restrictive & \tmatch{relay} & \tmatch{relay} \\
\bottomrule
\end{tabular}
\end{table}

Table~\ref{tab:classpairs} emulates all nine ordered pairs of NAT
behavior class against the deployed traversal implementation. The
reachability rule's prediction matched the outcome in seven pairs, shown
green, and mispredicted two, shown red; both errors forecast a relay where
traversal in fact succeeded directly, never the reverse. Both errors are the
ordinary-against-restrictive pairing, where an endpoint-independent
mapping lets the punch answer the source a packet arrived from rather
than the address that was advertised.

\subsection{Dividing the model}
\label{sec:eval-planner}

\begin{table}[t]
\caption{Kafila against baselines, best bold}
\label{tab:results}
\centering
\footnotesize
\setlength{\tabcolsep}{4pt}
\begin{tabular}{@{}l l r r r r@{}}
\toprule
Model & Allocation & Imbal. & Bneck & Util. & tok/s \\
      &            & \scriptsize max/min & \scriptsize ms & & \\
\midrule
\multicolumn{6}{@{}l}{\textbf{LAN}, 12\% network overhead}\\[1pt]
\multirow{3}{*}{Qwen3-8b} & Kafila & \textbf{1.46} & \textbf{11.3} & \textbf{86\%} & \textbf{21.5}\,{\scriptsize$\pm$0.2}\\
 & Mem-prop.\,(exo) & 4.01 & 26.0 & 48\% & 17.2\,{\scriptsize$\pm$3.8}\\
 & Uniform\,(GPipe) & 5.14 & 36.2 & 41\% & 13.7\,{\scriptsize$\pm$1.3}\\
\midrule
\multicolumn{6}{@{}l}{\textbf{US Central}, 38 to 62\% network overhead}\\[1pt]
\multirow{3}{*}{Qwen3-8b} & Kafila & \textbf{1.62} & \textbf{5.2} & \textbf{81\%} & \textbf{17.4}\,{\scriptsize$\pm$1.2}\\
 & Mem-prop.\,(exo) & 2.10 & 5.6 & 77\% & 15.0\,{\scriptsize$\pm$1.4}\\
 & Uniform\,(GPipe) & 1.66 & 5.6 & 81\% & 16.8\,{\scriptsize$\pm$2.1}\\
\cmidrule(l{2pt}){2-6}
\multirow{3}{*}{Qwen3-14b} & Kafila & \textbf{1.46} & \textbf{8.1} & \textbf{82\%} & \textbf{12.8}\,{\scriptsize$\pm$0.8}\\
 & Mem-prop.\,(exo) & 2.50 & 9.7 & 72\% & 12.4\,{\scriptsize$\pm$2.3}\\
 & Uniform\,(GPipe) & 2.07 & 9.6 & 75\% & 11.1\,{\scriptsize$\pm$1.2}\\
\cmidrule(l{2pt}){2-6}
\multirow{3}{*}{Qwen3-32b} & Kafila & \textbf{1.37} & \textbf{17.4} & \textbf{80\%} & 8.8\,{\scriptsize$\pm$1.4}\\
 & Mem-prop.\,(exo) & 2.83 & 21.2 & 69\% & \textbf{8.9}\,{\scriptsize$\pm$1.3}\\
 & Uniform\,(GPipe) & 2.28 & 20.8 & 74\% & 8.1\,{\scriptsize$\pm$1.1}\\
\midrule
\multicolumn{6}{@{}l}{\textbf{Intercontinental}, 84 to 94\% network overhead}\\[1pt]
\multirow{3}{*}{Qwen3-8b} & Kafila & \textbf{1.84} & \textbf{6.2} & \textbf{75\%} & \textbf{2.3}\,{\scriptsize$\pm$0.2}\\
 & Mem-prop.\,(exo) & 5.38 & 15.6 & 41\% & 2.3\,{\scriptsize$\pm$0.3}\\
 & Uniform\,(GPipe) & 10.11 & 24.9 & 33\% & 2.3\,{\scriptsize$\pm$0.2}\\
\cmidrule(l{2pt}){2-6}
\multirow{3}{*}{Qwen3-14b} & Kafila & \textbf{1.51} & \textbf{8.3} & \textbf{87\%} & \textbf{2.3}\,{\scriptsize$\pm$0.2}\\
 & Mem-prop.\,(exo) & 4.63 & 23.5 & 43\% & 2.2\,{\scriptsize$\pm$0.2}\\
 & Uniform\,(GPipe) & 10.59 & 43.5 & 32\% & 2.1\,{\scriptsize$\pm$0.1}\\
\cmidrule(l{2pt}){2-6}
\multirow{3}{*}{Qwen3-32b} & Kafila & \textbf{1.46} & \textbf{18.1} & \textbf{85\%} & \textbf{1.9}\,{\scriptsize$\pm$0.4}\\
 & Mem-prop.\,(exo) & 5.49 & 54.6 & 41\% & 1.9\,{\scriptsize$\pm$0.1}\\
 & Uniform\,(GPipe) & \multicolumn{4}{c}{\textit{infeasible}}\\
\bottomrule
\end{tabular}
\end{table}

Kafila divides against measured speed, and the effect on the division is
large and holds on every fleet (Table~\ref{tab:results}). Its slowest
stage exceeds its fastest by $1.4$ to $1.8\times$, where a memory-proportional
split reaches 2.1 to 5.5 and a uniform one up to 10.6. The heuristics do
not spread the excess evenly: each buries the laptop, which holds middling
memory and reads an order of magnitude slower than the fastest card, so a
share sized by memory is work it cannot get through, and in a ring every
token waits for it. Kafila's bottleneck stage is consequently up to $5.2\times$
shorter than uniform division and up to $3.0\times$ shorter than a
memory-proportional one. Read the other way this is utilisation, the
average stage's compute over the bottleneck's: the share of committed
device-time that computes rather than waiting on the slowest stage, which
Kafila holds at 75 to 87 per cent where the heuristics fall below half. It
is not the imbalance column renamed, which weighs only the fastest and
slowest stages; utilisation counts every one, and it sets the ceiling the
multi-user mode reaches: the four-user throughput gain climbs with it, from
near $2\times$ at the heuristics' 42 per cent to near $4\times$ at Kafila's
86 (Section~\ref{sec:eval-concurrency}). These are per-device times on each
device's own clock, independent of the network between them.

Getting the division right can decide whether a model runs at all. On the
intercontinental fleet no uniform split of Qwen3-32b fits the 18\,GB
laptop's equal share once cache is reserved for four concurrent requests,
so no session forms; the planner assigns an uneven division, names the
device a uniform split would overload, and serves the model. The same
memory wall bounds the LAN fleet's 12\,GB member on Qwen3-14b, so that
fleet is evaluated on Qwen3-8b.

What the equalized compute returns to a user depends on how much of a
token is computation rather than network, the overhead the fleets vary from
12 per cent to 94. Where members share a LAN the
planner's division delivers $1.56\times$ the throughput of a uniform
split and $1.25\times$ of a memory-proportional one, with non-overlapping
five-session ranges. The closer the members sit, the more of the
planner's compute advantage the user receives: as the network share
rises the same advantage occupies less of each token, so on the
intercontinental fleet the allocations serve at rates within the
five-session spread of one another even as their bottleneck and
utilisation stay as far apart as the table shows. The planner governs the
compute term, and a group whose devices share a network collects that
term in full.

The planner makes three decisions from one cost model, to different ends:
the division sets throughput, head placement keeps the plan within memory,
and the ring order holds network cost down. The throughput is the
division's, and the cleanest ablation of it is the headline: dividing evenly
or by memory, without the cost model, is the uniform and memory-proportional
split and forfeits the $5.2\times$ and $3.0\times$ of
Table~\ref{tab:results}. Pinning the head where the session opened, and
ordering by relay count rather than cost, then leave the served division,
its bottleneck, and its throughput within run-to-run variance on every
fleet. That is expected, since neither is a throughput knob and these fleets
do not stress it: the opener could hold the head, and every LAN and US edge
went direct. Their value is elsewhere, in a head that will not fit declining
a session and a cost-blind order relaying where it need not. The cost model
is fitted to nothing, yet predicts per-stage compute to a median of 18 per
cent across 165 stages, so the residual imbalance is that prediction gap,
not a limit of the method.

\begin{figure}[t]
\centering
\includegraphics[width=0.9\columnwidth]{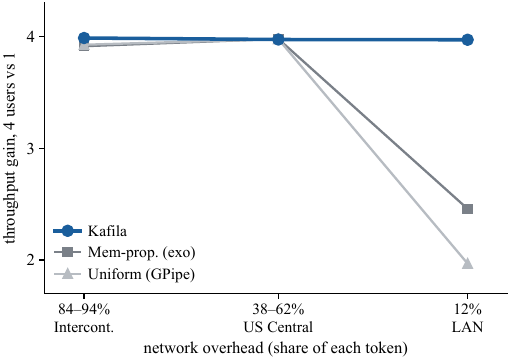}
\caption{Four-user throughput gain over one user against network overhead.
Only Kafila's balanced division scales near-linearly at every overhead; the
imbalanced divisions drop away as the overhead falls. Points are per-fleet
medians across models.}
\label{fig:concurrency}
\end{figure}

\subsection{Serving several users at once}
\label{sec:eval-concurrency}

Sequential decoding leaves a bubble at every stage: a token must pass
through all $k$ members before the next begins, so at any instant one
stage computes and the rest wait. Concurrent requests fill that bubble,
letting each stage work on one user's token while another stage works on
the next user's. The bubble is there at any network share; a slow network
only widens it. Figure~\ref{fig:concurrency} reports four concurrent users
against one on all three fleets; to check that the batch-one allocation is
still the right one under load, we run every division, not only the
planner's.

The LAN fleet is the decisive case, because at 12 per cent network
overhead there is almost no network idle to hide behind. There the planner's
division raises total throughput by $4.0\times$ while the memory-proportional
and uniform divisions reach only $2.5$ and $2.0\times$. The imbalanced
divisions stall at their overloaded stage, which was already near capacity
with one user, so extra requests queue behind it rather than filling other
members. Only the balanced division has the per-stage headroom to absorb
four requests, so the planner's single-user throughput lead of $1.6\times$
over uniform division widens to $3.2\times$ under four users. This answers
the concern that a bottleneck objective chosen for one request might be
wrong for many. Batching four requests raises arithmetic intensity, but
the member an imbalanced split overloads is the slowest on both axes that
shift, streaming weights and computing, so it stays the binding stage as
the regime moves; the division that unburdens it for one request is the
one that scales to four. Were the objective wrong for multi-user mode the
lead would shrink under load; on the LAN it doubles.

Where the network is slower it supplies the idle that a poor division
would otherwise lack, so every division fills and the gains converge. On
US Central all three reach $3.9$ to $4.0\times$ across the three models while
each user's rate changes by 1.0 per cent or less, and the planner leads on
absolute throughput at every size but the largest, where it ties. On the
intercontinental fleet, where network overhead is 84 to 94 per cent,
throughput scales $3.5$ to $4.0\times$ for all divisions and, as with a single
user at that distance, they fall within session spread: the network masks
the division. Serving four users is therefore near-free wherever a session
is worth forming, and where the members share a network the planner both
serves one user faster and admits more of them before its smallest device
runs out of cache.

% =========================================================================
\section{Conclusions and Future Work}
\label{sec:conclusion}

A trusted set of heterogeneous, commodity machines can serve an LLM that
none of them could run alone. A session must form
despite the NATs its members sit behind, and be worth using once formed. This paper proposed a protocol for the first
and a planner for the second, measured across three fleets that vary how
far apart the members sit. Every admitted session formed without router
configuration. Against the two divisions established in prior work, the
planner shortens the slowest stage by up to $5.2\times$ relative to uniform
division and up to $3.0\times$ relative to a memory-proportional one,
keeps 75 to 87 per cent of the committed hardware working where those
divisions fall below half, and serves a model uniform division cannot
place at all. Its cost model is fitted to nothing, yet predicts per-stage
compute to a median of 18 per cent. How much of that advantage a user
sees depends on how much of a token is computation rather than network
overhead, which we measured from 12 to 94 per cent: where members share a
network the same
division returns $1.56\times$ the throughput of uniform division and
$1.25\times$ of a memory-proportional one, and devices in one lab or home
sit at that end. Under four concurrent users that lead does not fade but
compounds to $3.2\times$, because a fast network leaves little idle for an
imbalanced division to hide behind, so the balance Kafila's planner strikes
is what lets a session scale to many users rather than merely serve one well.

Four directions remain for future work. The largest is the network term,
and these measurements bound what reducing it could recover. Membership that changes mid-session needs cache recovery around a replan the planner can already compute. A mobile member changes exactly the two properties the planner conditions on, so the question is how often to replan rather than how. And cache reserved for prompts that never arrive would return to the model if shared across members. A session of invited devices gives up the redundancy that lets a swarm
route around its weakest participant. What it offers in exchange is a
group whose membership, capability and connectivity are known before
serving begins, and which divides well enough to serve models no member
could hold alone.

% =========================================================================
\bibliographystyle{IEEEtran}
\bibliography{refs}

\end{document}